\documentclass[11pt,preprintnumbers,titlepage,nofootinbib]{revtex4-2}%
\usepackage[latin9]{inputenc}
\usepackage{float}
\usepackage{amsmath}
\usepackage{amssymb}
\usepackage{graphicx}
\usepackage[unicode=true,  bookmarks=false,  breaklinks=false,pdfborder={0 0
1},backref=section,colorlinks=false]{hyperref}
\usepackage{wasysym}
\usepackage{array}
\usepackage{multirow}
\usepackage{wasysym}
\usepackage{wasysym}
\usepackage{amsfonts}
\usepackage{subfigure}
\usepackage{cleveref}
\usepackage{listings}
\usepackage{xcolor}
\usepackage{array}
\usepackage{url}
\usepackage{color}%
\hypersetup{
colorlinks,linkcolor=blue,citecolor=blue,urlcolor=blue}
\makeatletter

\@ifundefined{textcolor}{}{
\definecolor{BLACK}{gray}{0}
\definecolor{WHITE}{gray}{1}
\definecolor{RED}{rgb}{1,0,0}
\definecolor{GREEN}{rgb}{0,1,0}
\definecolor{BLUE}{rgb}{0,0,1}
\definecolor{CYAN}{cmyk}{1,0,0,0}
\definecolor{MAGENTA}{cmyk}{0,1,0,0}
\definecolor{YELLOW}{cmyk}{0,0,1,0}
}

\makeatother
\begin{document}
\preprint{CTP-SCU/2024010}
\title{Spin-induced Scalar Clouds around Kerr-Newman Black Holes}
\author{Guangzhou Guo$^{b,a}$}
\email{guangzhouguo@outlook.com}
\author{Peng Wang$^{a}$}
\email{pengw@scu.edu.cn}
\author{Tianshu Wu$^{a}$}
\email{wutianshu@stu.scu.edu.cn}
\author{Haitang Yang$^{a}$}
\email{hyanga@scu.edu.cn}
\affiliation{$^{a}$Center for Theoretical Physics, College of Physics, Sichuan University,
Chengdu, 610064, China}
\affiliation{$^{b}$Department of Physics, Southern University of Science and Technology,
Shenzhen, 518055, China}

\begin{abstract}
Recent studies have demonstrated that a scalar field non-minimally coupled to
the electromagnetic field can experience a spin-induced tachyonic instability
near Kerr-Newman black holes, potentially driving the formation of scalar
clouds. In this paper, we construct such scalar clouds for both fundamental
and excited modes, detailing their existence domains and wave functions. Our
results indicate that a sufficiently strong coupling between the scalar and
electromagnetic fields is essential for sustaining scalar clouds. Within the
strong coupling regime, black holes that rotate either too slowly or too
rapidly are unable to support scalar clouds. Furthermore, we observe that
scalar cloud wave functions are concentrated near the black hole's poles.
These findings provide a foundation for future investigations of spin-induced
scalarized Kerr-Newman black holes.

\end{abstract}
\maketitle
\tableofcontents

\section{{}Introduction}

The no-hair theorem, which states that stationary black holes are uniquely
characterized by their mass, angular momentum and charge
\cite{Israel:1967wq,Carter:1971zc,Ruffini:1971bza,Wu:2024ixf,Chen:2023uuy}, is
a cornerstone of general relativity in the electro-vacuum context. Testing
this theorem is crucial for advancing our understanding of black hole physics
and for constraining the validity of alternative gravitational theories. For
example, the black-hole spectroscopy program, which analyzes quasinormal modes
extracted from gravitational-wave observations, has emerged as a valuable tool
for probing the Kerr nature of astrophysical compact objects
\cite{Isi:2019aib,Bhagwat:2019dtm,Wang:2021elt}.

Since the discovery of the first hairy black hole solution within
Einstein-Yang-Mills theory
\cite{Luckock:1986tr,Droz:1991cx,Kanti:1995vq,Sotiriou:2013qea,Cisterna:2014nua,Antoniou:2017acq}%
, numerous counterexamples to the no-hair theorem have appeared
\cite{Volkov:1989fi,Bizon:1990sr,Greene:1992fw}. In particular, black holes
with scalar hair have garnered significant attention due to the potential of
scalar fields to model dark energy and dark matter beyond the standard model
\cite{Herdeiro:2015waa}. A prime example is the presence of ultralight scalar
fields outside rotating black holes, which can undergo a superradiant
instability \cite{Brito:2015oca}, leading to the formation of scalar clouds
\cite{Hod:2014baa,Benone:2014ssa,Huang:2017whw,Kunz:2019bhm,Santos:2021nbf}.
The signatures of these scalar clouds have been used to impose stringent
constraints on the scalar field's parameter space, offering valuable insights
into dark matter exploration and beyond-the-Standard-Model physics
\cite{Arvanitaki:2009fg,Brito:2017wnc,Davoudiasl:2019nlo,Chen:2019fsq,Chen:2022kzv,Chen:2023vkq}%
. Moreover, under specific conditions, non-linear effects can evolve scalar
clouds into stationary hairy black holes
\cite{Herdeiro:2014goa,Delgado:2020hwr}.

Alternatively, non-minimal couplings between a scalar field and curvature
invariants have been shown to induce a tachyonic instability in the scalar
field \cite{Cardoso:2013opa,Cardoso:2013fwa}. This instability can lead to
spontaneous scalarization, endowing black holes with scalar hair only above a
certain threshold of spacetime curvature
\cite{Damour:1993hw,Doneva:2017bvd,Silva:2017uqg,Cunha:2019dwb,Cunha:2019dwb,Xu:2024cfe}%
. Consequently, spontaneous scalarization allows scalarized black holes to
acquire a non-trivial scalar configuration exclusively in regimes of strong
gravity, enabling them to evade constraints derived from weak-field gravity
tests. Moreover, it has been demonstrated that, within a specific parameter
region, the scalar field can exhibit a tachyonic instability near Kerr black
holes when the black hole's spin exceeds a certain threshold
\cite{Dima:2020yac}. Subsequently, this spin-induced tachyonic instability has
been shown to be capable of generating hairy (scalarized) black holes at
sufficiently high spins \cite{Herdeiro:2020wei,Berti:2020kgk}. In addition,
the dynamics of spontaneous scalarization around Kerr black holes has been
examined within the contexts of the Gauss-Bonnet and Chern-Simons gravities
\cite{Doneva:2021dqn,Doneva:2021dcc}. For a comprehensive overview of
spontaneous scalarization, we refer readers to the review presented in
\cite{Doneva:2022ewd}.

Similarly, in specific Einstein-Maxwell-scalar (EMS) models featuring a
non-minimal coupling between a scalar field and the Maxwell electromagnetic
invariant, the coupling, with an appropriate sign, can induce a tachyonic
instability in Reissner-Nordstr\"{o}m (RN) black holes \cite{Herdeiro:2018wub}%
. The evolutions of RN black holes into scalarized RN black holes have been
studied, providing valuable insights into spontaneous scalarization. Moreover,
for certain parameter regimes, scalarized RN black holes have been found to
possess two photon spheres outside the event horizon \cite{Gan:2021pwu}. This
unique feature leads to distinct phenomenology, including black hole images
with intricate structures
\cite{Gan:2021xdl,Guo:2022muy,Chen:2022qrw,Chen:2023qic,Chen:2024ilc} and echo
signals \cite{Guo:2021enm,Guo:2022umh}. Additionally, investigations into
superradiant instabilities and non-linear stability of these double photon
sphere black holes have been conducted \cite{Guo:2023ivz,Guo:2024cts}. For a
comprehensive analysis of black holes with multiple photon spheres, we refer
readers to \cite{Guo:2022ghl}.

Interestingly, the tachyonic instability persists even when RN black holes
rotate, leading to the formation of scalarized Kerr-Newman (KN) black holes
\cite{Guo:2023mda}. The existence of these black holes is bounded by
bifurcation points, corresponding to scalar clouds supported by KN black
holes. However, the presence of scalarized KN black holes is suppressed by the
black hole's spin, with a maximum spin threshold beyond which such solutions
cease to exist. An analysis of scalar clouds induced by the tachyonic
instability around KN black holes has also been conducted, revealing that
black holes with sufficiently large spin cannot support scalar clouds
\cite{Guo:2024bkw}. Conversely, if the sign of the coupling constant is
reversed, a spin-induced tachyonic instability emerges in KN black holes when
they rotate sufficiently fast \cite{Hod:2022txa,Lai:2022ppn}. This tachyonic
instability can trigger the formation of scalar clouds around KN black holes,
marking the onset of spin-induced scalarized KN black holes. The existence
domain of spin-induced scalar clouds has been investigated only for a limited
range of black hole parameters, particularly in the strong coupling limit.
Therefore, a more comprehensive exploration of spin-induced scalar clouds is
necessary for a deeper understanding of spontaneous scalarization in KN black
holes, providing a foundation for constructing spin-induced scalarized KN
black holes.

This paper presents a comprehensive investigation of scalar clouds generated
by the spin-induced tachyonic instability within the EMS model. The paper is
organized as follows. In Sec. \ref{sec:Setup}, we introduce the EMS model and
its associated scalar clouds, followed by an overview of the computational
framework employed to obtain scalar clouds using the spectral method. Sec.
\ref{Sec:Conc} presents and analyzes our numerical findings. Finally, Section
\ref{Sec:Conc} summarizes key results and discusses their implications.
Throughout this paper, we adopt units where $G=c=4\pi\epsilon_{0}=1$.

\section{Setup}

\label{sec:Setup}

This section commences with a review of the EMS model and the conditions under
which it exhibits a spin-induced tachyonic instability. Subsequently, we
investigate the scalar clouds generated by this tachyonic instability and
present the numerical method for determining their existence domains and wave functions.

\subsection{Tachyonic Instability}

A tachyonic instability emerges within the EMS model, where a scalar field
$\Phi$ is non-minimally coupled to the electromagnetic field $A_{\mu}$ through
a coupling function $f\left(  \Phi\right)  $. Explicitly, the action is given
by
\begin{equation}
S=\frac{1}{16\pi}\int d^{4}x\sqrt{-g}\left[  R-2\partial_{\mu}\Phi
\partial^{\mu}\Phi-f\left(  \Phi\right)  F^{\mu\nu}F_{\mu\nu}\right]  ,
\label{eq:Action}%
\end{equation}
where $F_{\mu\nu}=\partial_{\mu}A_{\nu}-\partial_{\nu}A_{\mu}$ represents the
electromagnetic field strength tensor. Varying the action yields the equation
of motion for $\Phi$,
\begin{equation}
\square\Phi=f^{\prime}\left(  \Phi\right)  F_{\mu\nu}F^{\mu\nu}/4.
\label{eq:seq}%
\end{equation}
Remarkably, the inclusion of the scalar-electromagnetic non-minimal coupling
term induces the tachyonic instability in the scalar field, leading to
spontaneous scalarization in black holes \cite{Herdeiro:2018wub,Guo:2023mda}.
For spontaneous scalarization to occur, a scalar-free solution with $\Phi=0$
must exist, from which scalar hair can develop. This condition imposes
$f^{\prime}\left(  0\right)  \equiv\left.  df\left(  \Phi\right)
/d\Phi\right\vert _{\Phi=0}=0$, resulting in the series expansion of $f\left(
\Phi\right)  $ around $\Phi=0$,%
\begin{equation}
f\left(  \Phi\right)  =1+\alpha\Phi^{2}+\mathcal{O}\left(  \Phi^{3}\right)  ,
\end{equation}
where $\alpha$ is a dimensionless coupling constant quantifying the strength
of the scalar-electromagnetic interaction. Without loss of generality, we set
$f\left(  0\right)  =1$.

Within the EMS model, the rotating scalar-free black hole solution is a KN
black hole. Expressed in Boyer-Lindquist coordinates, its metric and vector
potential are given by%

\begin{align}
ds^{2}  &  =-\frac{\triangle}{\Sigma}\left(  dt-a\text{sin}^{2}\theta
d\varphi\right)  ^{2}+\frac{\text{sin}^{2}\theta}{\Sigma}\left[  \left(
r^{2}+a^{2}\right)  d\varphi-adt\right]  ^{2}+\frac{\Sigma}{\triangle}%
dr^{2}+\Sigma d\theta^{2},\nonumber\\
A  &  =Qr\frac{dt-a\text{sin}^{2}\theta d\varphi}{\Sigma},
\end{align}
where\textbf{ }
\begin{align}
\Sigma &  =r^{2}+a^{2}\cos^{2}\theta,\nonumber\\
\triangle &  =r^{2}-2Mr+a^{2}+Q^{2}.
\end{align}
Here, $Q$ is the black hole charge, and $a$ represents the ratio of black hole
angular momentum $J$ to mass $M$ (i.e., $a\equiv J/M$). The event and Cauchy
horizons are located at the roots of\textbf{ $\triangle$, }given by
$r_{+}=M+\sqrt{M^{2}-a^{2}-Q^{2}}$ and $r_{-}=M-\sqrt{M^{2}-a^{2}-Q^{2}}$,
respectively. For convenience, we introduce the dimensionless reduced black
hole charge and spin, defined as
\begin{equation}
q\equiv Q/M\text{, }\chi\equiv a/M\text{.}%
\end{equation}

To investigate the stability of the scalar field in the scalar-free black hole
background, we adopt the probe limit, neglecting the scalar field's
backreaction. Within this approximation, the scalar field obeys the equation
of motion
\begin{equation}
\left(  \square-\mu_{\text{eff}}^{2}\right)  \Phi=0, \label{eq:phiEq}%
\end{equation}
where $\mu_{\text{eff}}^{2}=$ $\alpha F_{\mu\nu}F^{\mu\nu}/2$ is the effective
mass squared. Self-interactions of the scalar field, which have a minimal
impact on the onset of spontaneous scalarization
\cite{Herdeiro:2018wub,Fernandes:2019rez,Hod:2020ljo}, are disregarded. For KN
black holes, the effective mass squared becomes
\begin{equation}
\mu_{\text{eff}}^{2}=-\frac{\alpha q^{2}\left(  \tilde{r}^{4}-6\chi^{2}%
\tilde{r}^{2}\cos^{2}\theta+\chi^{4}\cos^{4}\theta\right)  }{\left(  \tilde
{r}^{2}+\chi^{2}\cos^{2}\theta\right)  ^{4}M^{2}},
\end{equation}
where $\tilde{r}\equiv r/M$. Given the spatial dependence of $\mu_{\text{eff}%
}^{2}$, the tachyonic instability is indicated by%
\begin{equation}
\min\mu_{\text{eff}}^{2}<0.
\end{equation}
Moreover, when the tachyonic instability arises, the minimum value of
$\mu_{\text{eff}}^{2}$ becomes increasingly negative for a fixed $\chi$ as the
magnitude of $\alpha$ or $q$ increases, signifying an amplification of the
tachyonic instability with larger $\left\vert \alpha\right\vert $ or $q$ values.

The occurrence of the tachyonic instability has been explored for both
$\alpha>0$ \cite{Guo:2023mda} and $\alpha<0$ \cite{Hod:2022txa,Lai:2022ppn}.
In the case of $\alpha>0$, the region where $\mu_{\text{eff}}^{2}<0$ has been
shown to exist outside the event horizon of KN black holes. The spatial extent
of this region diminishes with increasing black hole spin, suggesting a
potential suppression of the tachyonic instability for rapidly rotating black
holes. For $\alpha<0$, both spin $\chi$ and charge $q$ must be non-zero for
the tachyonic instability to arise. Specifically, the parameter space
admitting the tachyonic instability is constrained by%
\begin{equation}
\chi\geq\frac{1+\sqrt{1-2\left(  2-\sqrt{2}\right)  q^{2}}}{2\sqrt{2}}\text{
with }0<q\leq q_{\text{cr}}\equiv\sqrt{2\sqrt{2}-2}\text{.} \label{eq:consxq}%
\end{equation}
The global minimum value of $\chi$ imposes a lower bound,%
\begin{equation}
\chi\geq\chi_{\text{cr}}\equiv\sqrt{2}-1. \label{eq:constx}%
\end{equation}
It is noteworthy that $\chi_{\text{cr}}^{2}+q_{\text{cr}}^{2}=1$, indicating
that the KN black hole with $q=q_{\text{cr}}$ and $\chi=\chi_{\text{cr}}$ is extremal.

\subsection{Scalar Clouds}

The regular bound-state solutions of Eq. $\left(  \ref{eq:phiEq}\right)  $ are
interpreted as scalar clouds surrounding KN black holes. The tachyonic
instability can serve as a driving mechanism for the formation of scalar
clouds. Indeed, scalar clouds around KN black holes induced by the tachyonic
instability have been recently investigated for the $\alpha>0$ case
\cite{Guo:2024bkw}. This paper focuses on scalar clouds in the $\alpha<0$
regime. As the formation of such scalar clouds necessitates a tachyonic
instability, the bounds on $\chi$ and $q$ imposed by Eqs. $\left(
\ref{eq:consxq}\right)  $ and $\left(  \ref{eq:constx}\right)  $ constrain
their existence domain in the parameter space.

Leveraging the axial symmetry of KN black holes, we decompose the scalar field
$\Phi$ into a Fourier series in terms of frequency $\omega$ and azimuthal
number $m$,%
\begin{equation}
\Phi\left(  t,r,\theta,\varphi\right)  =\int\frac{d\omega}{2\pi}e^{-i\omega
t}\sum\limits_{m}e^{im\varphi}\tilde{\Phi}\left(  \omega,r,\theta,m\right)  .
\end{equation}
For specified $\omega$ and $m$, Eq. $\left(  \ref{eq:phiEq}\right)  $ reduces
to a Partial Differential Equation (PDE) for $\tilde{\Phi}\left(
\omega,r,\theta,m\right)  $ with respect to $r$ and $\theta$. As scalar clouds
typically serve as seeds for constructing axisymmetric hairy black hole
solutions, we focus on stationary, axisymmetric configurations by setting
$\omega=m=0$. For brevity, we denote $\tilde{\Phi}\left(  0,r,\theta,0\right)
$ by $\phi\left(  r,\theta\right)  $ in subsequent discussions. Analogous to
hydrogen atoms, wave functions $\phi\left(  r,\theta\right)  $ can be
characterized by a discrete set of numbers $\left(  n,l\right)  $, where the
principal quantum number $n=0,1,2\cdots$ and the angular momentum quantum
number $l=0,1,2\cdots$ correspond to the number of nodes of wave functions in
the radial and angular directions, respectively.

To determine $\phi\left(  r,\theta\right)  $, appropriate boundary conditions
must be imposed at the event horizon and spatial infinity. Given the
regularity of $\phi\left(  r,\theta\right)  $ across the event horizon, it can
be expanded in a series about $r=r_{+}$,%
\begin{equation}
\phi\left(  r,\theta\right)  =\phi_{0}\left(  \theta\right)  +\left(
r-r_{+}\right)  \phi_{1}\left(  \theta\right)  +\cdots. \label{eq:phiexp}%
\end{equation}
Moreover, the condition of asymptotic flatness necessitates $\phi\left(
r,\theta\right)  $ vanishing as $r$ approaches infinity,%
\begin{equation}
\lim_{r\rightarrow\infty}\phi\left(  r,\theta\right)  =0\text{.}%
\end{equation}
Additionally, axial symmetry, coupled with regularity on the symmetry axis,
enforces,%
\begin{equation}
\partial_{\theta}\phi\left(  r,\theta\right)  =0\text{, at }\theta=0\text{ and
}\pi\text{.}%
\end{equation}
These boundary conditions uniquely select a discrete set of KN black holes
capable of supporting scalar clouds, thereby defining existence surfaces in
the $\left(  \alpha,\chi,q\right)  $ parameter space and existence lines
within these surfaces for fixed $\alpha$ in the $\left(  \chi,q\right)  $ plane.

As demonstrated in the $\alpha>0$ case, the existence lines of scalar clouds
in the $\left(  \chi,q\right)  $ parameter space delineate boundaries between
regions exhibiting excessively strong tachyonic instability for stationary
scalar clouds and those with insufficient instability for their formation
\cite{Guo:2023mda,Guo:2024bkw}. For KN black holes constrained by Eqs.
$\left(  \ref{eq:consxq}\right)  $ and $\left(  \ref{eq:constx}\right)  $,
their minimum value of $\mu_{\text{eff}}^{2}$ approaches $-\infty$ as
$\alpha\rightarrow-\infty$. This observation suggests that, in the limit of
$\alpha\rightarrow-\infty$, parameter regions defined by $\left(
\ref{eq:consxq}\right)  $ and $\left(  \ref{eq:constx}\right)  $ may exhibit
excessively strong tachyonic instability for stationary scalar clouds.
Conversely, $\mu_{\text{eff}}^{2}$ of black holes outside these constrained
regions is always positive, indicating the absence of the tachyonic
instability. Consequently, the existence lines of scalar clouds coincide with
the boundaries of the constrained parameter regions. Specifically, as
$\alpha\rightarrow-\infty$, the existence lines in the $\left(  \chi,q\right)
$ parameter space converge to a critical existence line, given by%
\begin{equation}
\chi=\frac{1+\sqrt{1-2\left(  2-\sqrt{2}\right)  q^{2}}}{2\sqrt{2}}\text{ for
}0<q\leq q_{\text{cr}}\text{.} \label{eq:criticalLine}%
\end{equation}
Moreover, as $\chi$ increases, the critical existence line extends from
$\left(  \chi,q\right)  =\left(  \chi_{\text{cr}},q_{\text{cr}}\right)  $ to
$\left(  \chi,q\right)  =\left(  1/\sqrt{2},0\right)  $.

\subsection{Numerical Scheme}

The wave equation governing $\phi\left(  r,\theta\right)  $ in KN black holes
is separable, enabling its reduction to ordinary differential equations.
However, this study employs a spectral method to directly solve the wave
equation for $\phi\left(  r,\theta\right)  $, circumventing the need for
separability. Consequently, this approach offers a significant advantage for
computing scalar clouds around black holes in frameworks beyond general
relativity. Spectral methods, a well-established method for solving PDEs
\cite{boyd2001chebyshev}, approximate the exact solution through a finite
linear combination of basis functions. Notably, they exhibit exponential
convergence for well-behaved functions, surpassing the linear or polynomial
convergence rates achieved by finite difference and finite element methods.
Recent investigations have successfully applied spectral methods to the
identification of scalar cloud configurations \cite{Guo:2024bkw}, the
construction of black hole solutions
\cite{Fernandes:2022gde,Lai:2023gwe,Burrage:2023zvk} and the calculation of
black hole quasinormal modes
\cite{Jansen:2017oag,Gan:2019jac,Chung:2023zdq,Chung:2023wkd,Chung:2024ira}. A
comprehensive overview of spectral methods in this context can be found in
\cite{Fernandes:2022gde}.

To facilitate numerical implementation, we introduce a compact radial
coordinate defined as
\begin{equation}
x=\frac{\sqrt{r^{2}-r_{+}^{2}}-r_{+}}{\sqrt{r^{2}-r_{+}^{2}}+r_{+}},
\end{equation}
which maps the event horizon and spatial infinity to $x=-1$ and $x=1$,
respectively. Under this transformation, the boundary conditions at the event
horizon and spatial infinity become%
\begin{equation}
\partial_{x}\phi\left(  x,\theta\right)  =0\text{ and }\phi\left(
1,\theta\right)  =0,
\end{equation}
respectively. Without loss of generality, we assume that wave functions
$\phi\left(  x,\theta\right)  $ possess definite parity with respect to the
equatorial plane, thereby permitting the restriction of the analysis to the
upper half-domain $0\leq\theta\leq\pi/2$. For even and odd parities, the
boundary condition at $\theta=\pi/2$ is $\partial_{\theta}\phi\left(
x,\theta\right)  =0$ and $\phi\left(  x,\theta\right)  =0$, respectively. At
$\theta=0$, we have $\partial_{\theta}\phi\left(  x,\theta\right)  =0$.

To apply the spectral method, the function $\phi\left(  x,\theta\right)  $ is
decomposed into a spectral expansion as
\begin{equation}
\phi\left(  x,\theta\right)  =%
{\displaystyle\sum\limits_{i=0}^{N_{x}-1}}
{\displaystyle\sum\limits_{j=0}^{N_{\theta}-1}}
\alpha_{ij}T_{i}\left(  x\right)  \Theta_{j}\left(  \theta\right)  ,
\label{eq:sexpansion}%
\end{equation}
where $N_{x}$ and $N_{\theta}$ denote the resolutions in the radial and
angular coordinates, respectively, $T_{i}\left(  x\right)  $ represents the
Chebyshev polynomial, and $\alpha_{ij}$ are the spectral coefficients. The
angular basis $\Theta_{j}\left(  \theta\right)  $ is dependent on the parity
with respect to $\theta=\pi/2$. Specifically, we adopt
\begin{equation}
\Theta_{j}\left(  \theta\right)  =\left\{
\begin{array}
[c]{c}%
\cos\left(  2j\theta\right)  \text{ \ for even parity }\\
\cos\left[  \left(  2j+1\right)  \theta\right]  \text{ for odd parity}%
\end{array}
\right.  \text{.} \label{eq:sexpansion-1-1}%
\end{equation}
This choice ensures that $\phi\left(  x,\theta\right)  $ automatically
satisfies the boundary conditions at $\theta=0$ and $\pi/2$.

To determine the spectral coefficients $\alpha_{ij}$, the spectral expansion
$\left(  \ref{eq:sexpansion}\right)  $ is substituted into the PDE, followed
by discretization at the Gauss-Chebyshev points. This procedure transforms the
PDE for $\phi\left(  x,\theta\right)  $ into a system of algebraic equations
involving $\alpha_{ij}$. However, Eq. $\left(  \ref{eq:phiEq}\right)  $
exhibits linear scaling invariance, necessitating an additional constraint to
guarantee a non-trivial solution for $\alpha_{ij}$. This is achieved by
setting $\phi\left(  x,\theta\right)  =1$ at $\left(  x,\theta\right)
=\left(  -1,0\right)  $. This constraint introduces an extra algebraic
equation for $\alpha_{ij}$ through the spectral expansion $\left(
\ref{eq:sexpansion}\right)  $. To balance the number of unknowns and
equations, one black hole parameter (e.g., the reduced black hole charge $q$)
is treated as an additional unknown. The resulting system of algebraic
equations for $\alpha_{ij}$ and $q$ is then solved iteratively using the
Newton-Raphson method. At each iteration, the linear system of equations is
solved using the built-in LinearSolve command in Mathematica. The
Newton-Raphson algorithm iterates until successive iterations converge to
within a tolerance of $10^{-10}$. Moreover, while exploring scalar cloud
solutions within the $\left(  \alpha,\chi,q\right)  $ parameter space, the
residual of the spectral approximation and the number of nodes are monitored
to ensure solution accuracy, maintaining a residual tolerance of $10^{-7}$.

In the Appendix, we perform a convergence test of scalar cloud solutions by
plotting the residual error as a function of $N_{x}$ and $N_{\theta}$. The
results demonstrate that the error decays exponentially until reaching a
round-off plateau below $10^{-7}$. To balance numerical precision and
efficiency, we employ $\left(  N_{x},N_{\theta}\right)  =\left(  28,5\right)
$ for subsequent numerical computations of $\phi\left(  x,\theta\right)  $.

\section{Results}

\label{sec:NR}

In this section, we present numerical results concerning the parameter space
of KN black holes that can support scalar clouds for the fundamental and first
two excited modes. We also provide representative examples of the
corresponding scalar cloud wave functions.\begin{figure}[ptb]
\begin{centering}
\includegraphics[scale=0.75]{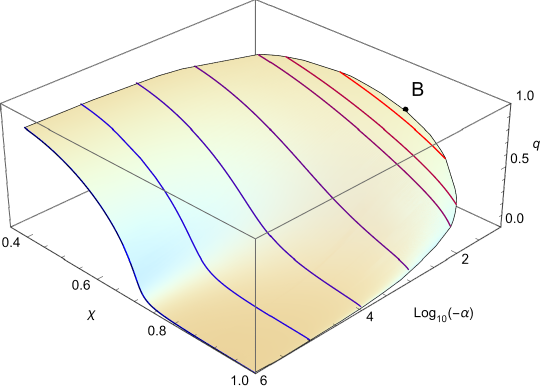}\includegraphics[scale=0.72]{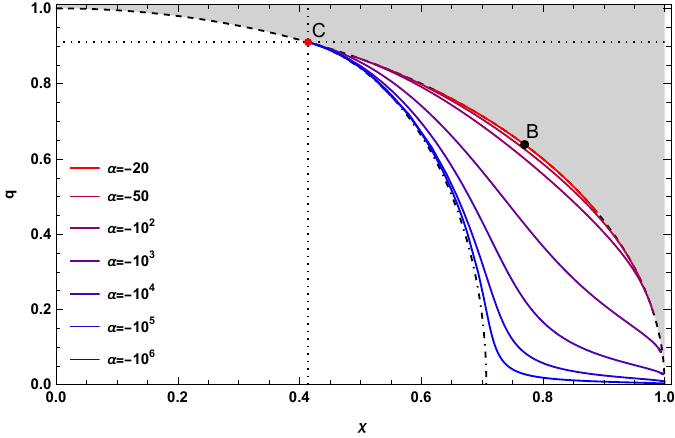}
\par\end{centering}
\caption{\textbf{Left Panel:} Existence surface of fundamental scalar clouds
with $\left(  n,l\right)  =\left(  0,0\right)  $ in the $\left(  \alpha
,\chi,q\right)  $ parameter space. KN black holes residing on the colored
surface admit the scalar clouds. The existence lines for $\alpha=-20$, $-50$,
$-10^{2}$, $-10^{3}$, $-10^{4}$, $-10^{5}$ and $-10^{6}$ are shown from right
to left. As $\alpha$ increases, this existence surface gradually converges to
the critical point $B$ with $\alpha=\alpha_{\text{cr}}\simeq-13.398$. Scalar
clouds cease to exist when $\alpha>\alpha_{\text{cr}}$. \textbf{Right Panel:}
Existence lines for $\alpha=-20$, $-50$, $-10^{2}$, $-10^{3}$, $-10^{4}$,
$-10^{5}$ and $-10^{6}$ in the $\left(  \chi,q\right)  $ space, displayed from
top right to bottom left. Both endpoints of the existence lines lie on the
extremal KN black hole line (black dashed line), beyond which KN black holes
cannot exist (gray region). As $\alpha$ decreases from $\alpha_{\text{cr}}$,
the existence line emerges from the critical point $B$ and gradually stretches
out. In the limit of $\alpha\rightarrow-\infty$, the left segment of the
existence line approaches the critical existence line (black dot-dashed line)
while the right segment approaches the $\chi$-axis with $q=0$. The vertical
and horizontal black dotted lines represent $\chi=\chi_{\text{cr}}$ and
$q=q_{\text{cr}}$, respectively, with their intersection marking the critical
point $C$. As $\alpha\rightarrow-\infty$, the left and right endpoints of
existence lines move along the extremal line towards the critical point $C$
and the point at $\left(  \chi,q\right)  =\left(  1,0\right)  $,
respectively.}%
\label{fig:ExistenceDomainZeroMode}%
\end{figure}

We begin by analyzing the fundamental mode of scalar clouds, characterized by
nodeless wave functions with $\left(  n,l\right)  =\left(  0,0\right)  $. The
left panel of Fig. \ref{fig:ExistenceDomainZeroMode} displays the existence
domain for fundamental clouds within the $\left(  \alpha,\chi,q\right)  $
parameter space. KN black holes supporting fundamental clouds reside on the
colored surface, while existence lines for various fixed $\alpha$ values are
also shown. Our findings reveal that as $\alpha$ increases, these existence
lines contract and converge towards the critical point $B$ at $\left(
\alpha,\chi,q\right)  \simeq\left(  -13.398,0.77001,0.63803\right)  $,
indicated by a black dot. Consequently, there exists a critical value of
$\alpha$, $\alpha_{\text{cr}}\simeq-13.398$, beyond which the spin-induced
tachyonic instability is insufficient to form scalar clouds. The right panel
of Fig. \ref{fig:ExistenceDomainZeroMode} illustrates the same existence lines
in the $\left(  \chi,q\right)  $ plane. The extremal KN black hole line,
corresponding to the condition $q^{2}+\chi^{2}=1$, is represented by a black
dashed line. KN black holes cannot exist in the gray region above this
extremal line, imposing an upper limit on the black hole charge $q$ for a
given $\chi$. The vertical and horizontal dotted lines correspond to
$\chi=\chi_{\text{cr}}$ and $q=q_{\text{cr}}$, respectively. The intersection
of these two dotted lines determines the critical point $C$ at $\left(
\chi,q\right)  =\left(  \chi_{\text{cr}},q_{\text{cr}}\right)  $, marked by a
red dot, which lies on the extremal line. Additionally, the black dot-dashed
line represents the critical existence line given by Eq. $\left(
\ref{eq:criticalLine}\right)  $.

Four key characteristics are observed regarding the existence lines:

\begin{itemize}
\item \textbf{Termination on Extremal Line:} Both endpoints of existence lines
lie on the extremal line. We assume that the left and right endpoints of the
existence line with a given $\alpha$ locate at $\left(  \chi,q\right)
=\left(  \chi_{\text{low}}\left(  \alpha\right)  ,q_{\text{up}}\left(
\alpha\right)  \right)  $ and $\left(  \chi,q\right)  =\left(  \chi
_{\text{up}}\left(  \alpha\right)  ,q_{\text{low}}\left(  \alpha\right)
\right)  $, respectively. The existence line decreases as $\chi$ increases,
implying $q_{\text{low}}\left(  \alpha\right)  <q_{\text{up}}\left(
\alpha\right)  $ and $\chi_{\text{low}}\left(  \alpha\right)  <\chi
_{\text{up}}\left(  \alpha\right)  $. As $\chi$ approaches $\chi_{\text{low}%
}\left(  \alpha\right)  $ from the right or $\chi_{\text{up}}\left(
\alpha\right)  $ from the left, the charge of the existence line converges to
the upper limit set by the extremal line. This implies that, when $\chi
<\chi_{\text{low}}\left(  \alpha\right)  $ or $\chi>\chi_{\text{up}}\left(
\alpha\right)  $, the presence of scalar clouds would necessitate a black hole
charge $q$ exceeding its extremal limit. Consequently, scalar clouds cease to
exist if $\chi<\chi_{\text{low}}\left(  \alpha\right)  $ or $\chi
>\chi_{\text{up}}\left(  \alpha\right)  $, due to insufficient tachyonic instability.

\item \textbf{Shift Toward Smaller }$q$\textbf{ with Decreasing }$\alpha
$\textbf{:} For $\alpha=\alpha_{\text{cr}}$, the existence line shrinks to the
critical point $B$ with\ $\chi_{\text{low}}\left(  \alpha_{\text{cr}}\right)
=\chi_{\text{up}}\left(  \alpha_{\text{cr}}\right)  \simeq0.77001$ and
$q_{\text{low}}\left(  \alpha_{\text{cr}}\right)  =q_{\text{up}}\left(
\alpha_{\text{cr}}\right)  \simeq0.63803$. As $\alpha$ decreases from
$\alpha_{\text{cr}}$, the existence lines shift towards smaller $q$ values
with increasing length. This is attributed to the enhancement of the tachyonic
instability for more negative $\alpha$, thereby permitting a lower $q$ to
support scalar cloud formation.

\item \textbf{Approach Critical Existence Line as }$\alpha\rightarrow-\infty$:
As $\alpha$ goes to $-\infty$, the segment of the existence line with
$\chi_{\text{low}}\left(  \alpha\right)  \leq\chi\leq1/\sqrt{2}$ converges to
the critical existence line, consistent with the preceding discussion.
Additionally, our results demonstrate that the segment with $1/\sqrt{2}%
<\chi\leq\chi_{\text{up}}\left(  \alpha\right)  $ approaches the $q=0$ line as
$\alpha\rightarrow-\infty$. This observation indicates that, when the coupling
constant $\alpha$ is sufficiently strong, a tiny amount of charge can trigger
the formation of scalar clouds if $\chi$ exceeds $1/\sqrt{2}$. Moreover, in
this limit, the left endpoint of the existence line approaches the critical
point $C$ while the right endpoint approaches the point at $\left(
\chi,q\right)  =\left(  1,0\right)  $.

\item \textbf{Serve as Threshold Line}: For a given $\alpha$, the existence
line can be considered a threshold line, below which KN black holes exhibit
insufficient tachyonic instability to support scalar clouds due to their low
$q$ values. Conversely, KN black holes above the existence line possess
excessively strong tachyonic instability to sustain stationary scalar clouds.
Therefore, non-linear effects are required to suppress this instability,
potentially giving rise to stationary states, such as scalarized KN black holes.
\end{itemize}

\begin{figure}[ptb]
\begin{centering}
\includegraphics[scale=0.555]{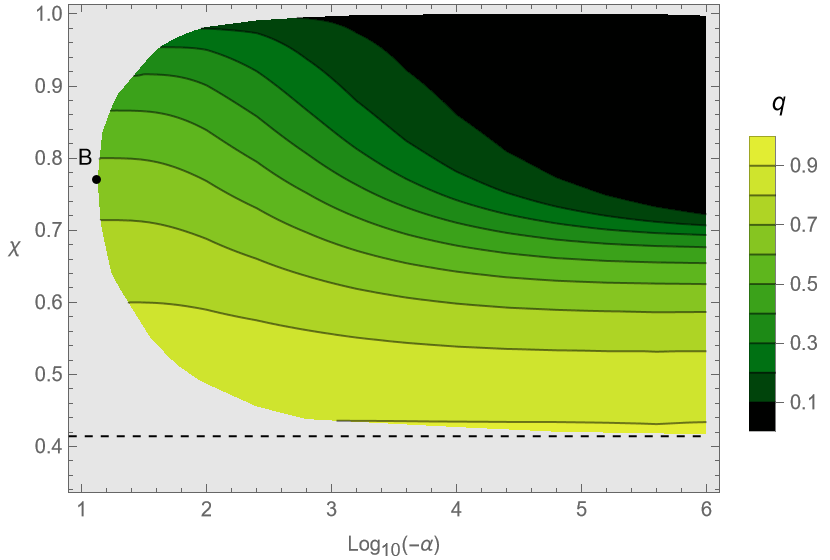} \includegraphics[scale=0.555]{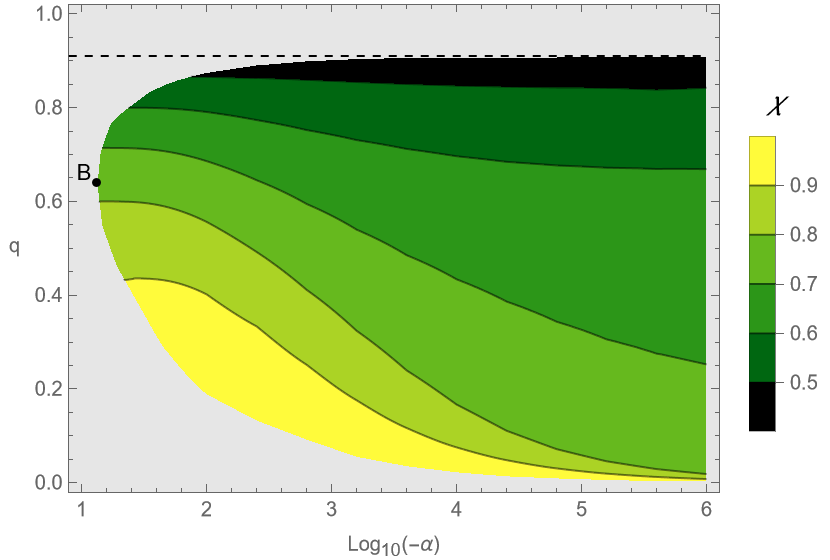}
\par\end{centering}
\caption{\textbf{Left Panel: }Existence domain of fundamental scalar clouds in
the $\left(  \alpha,\chi\right)  $ plane, represented by a density plot where
colors indicate the values of $q$. The domain is confined by the upper and
lower boundaries, defined by the right and left endpoints of the existence
lines, respectively. These boundaries merge at the critical point $B$,
indicating the upper limit $\alpha_{\text{cr}}$ on the coupling constant
required to support scalar clouds. The horizontal dashed line represents
$\chi=\chi_{\text{cr}}$, above which the existence domain is located.
\textbf{Right Panel:} Existence domain in the $\left(  \alpha,q\right)  $
plane, shown as a density plot with colors corresponding to $\chi$ values. The
domain is bounded by the upper and lower limits, formed by the left and right
endpoints of the existence lines, respectively. The horizontal dashed line
depicts $q=q_{\text{cr}}$, below which the existence domain is located.}%
\label{fig:densityplot}%
\end{figure}

The left and right panels of Fig. \ref{fig:densityplot} present density plots
of the fundamental cloud existence domain in the $\left(  \alpha,\chi\right)
$ and $\left(  \alpha,q\right)  $ spaces, respectively. Both panels illustrate
the absence of scalar cloud solutions for $\alpha>\alpha_{\text{cr}}$, where
$\alpha_{\text{cr}}$ is the $\alpha$ value of the critical point $B$. In the
$\left(  \alpha,\chi\right)  $ plane, the density plot colors represent the
magnitude of $q$, with grey regions indicating the non-existence of scalar
clouds. The existence domain is bounded by the upper limit line,
$\chi_{\text{up}}\left(  \alpha\right)  $, and the lower limit line,
$\chi_{\text{lower}}\left(  \alpha\right)  $. As $\alpha$ increases towards
$\alpha_{\text{cr}}$, the region of existence for scalar clouds contracts,
ultimately converging at the critical point $B$. In the limit of
$\alpha\rightarrow-\infty$, the upper and lower limits approach $1$ and
$\chi_{\text{cr}}$, respectively. Similarly, the existence domain in the
$\left(  \alpha,q\right)  $ plane is confined by the upper boundary,
$q_{\text{up}}\left(  \alpha\right)  $, and the lower boundary,
$q_{\text{lower}}\left(  \alpha\right)  $, which merge at the critical point
$B$. In the limit of $\alpha\rightarrow-\infty$, the upper and lower
boundaries approach $q_{\text{cr}}$ and $0$, respectively.

\begin{figure}[ptb]
\begin{centering}
\includegraphics[scale=0.73]{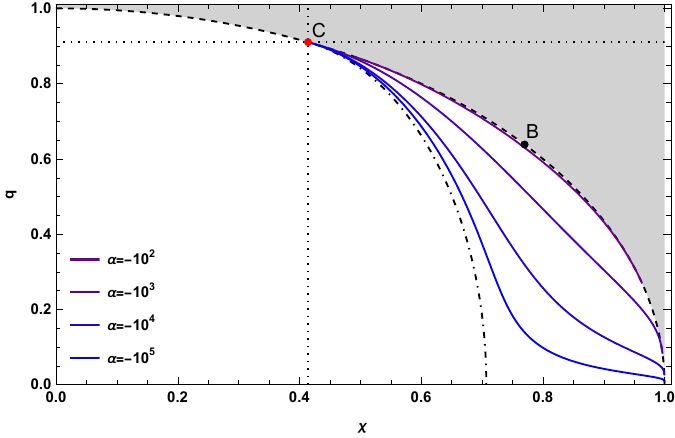}\includegraphics[scale=0.73]{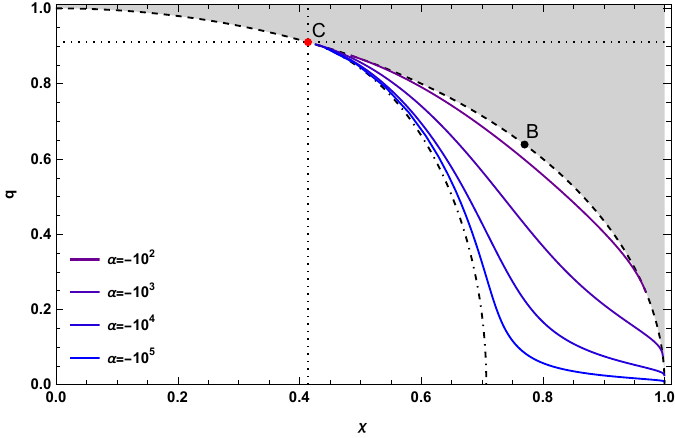}
\par\end{centering}
\caption{Existence lines in the $\left(  \chi,q\right)  $ space for excited
scalar clouds with $\left(  n,l\right)  =\left(  1,0\right)  $ (\textbf{Left
Panel}) and $\left(  n,l\right)  =\left(  0,1\right)  $ (\textbf{Right
Panel}). From top right to bottom left, the coupling constant $\alpha$ takes
the values of $-10^{2}$, $-10^{3}$, $-10^{4}$ and $-10^{5}$, respectively.
These existence lines closely resemble those of the fundamental scalar clouds.
Representative $q$ and $\chi$ values of the existence lines are listed in Tab.
\ref{tab:qanx}. It is evident that the existence lines for $\left(
n,l\right)  =\left(  1,0\right)  $ lie above those for $\left(  n,l\right)
=\left(  0,1\right)  $, suggesting that excited scalar clouds with $\left(
n,l\right)  =\left(  1,0\right)  $ require a stronger tachyonic instability.}%
\label{fig:ExistenceDomainExitedMode}%
\end{figure}

\begin{table}[ptb]%
\begin{tabular}
[c]{c|ccc|ccc|ccc|ccc}\hline
& \multicolumn{12}{|c}{$q$}\\\cline{2-13}\cline{5-7}%
$\chi$ & \multicolumn{3}{|c|}{$\alpha=-10^{2}$} & \multicolumn{3}{|c|}{$\alpha
=-10^{3}$} & \multicolumn{3}{|c|}{$\alpha=-10^{4}$} &
\multicolumn{3}{|c}{$\alpha=-10^{5}$}\\\cline{2-13}\cline{5-7}
& $(0,0)$ & $(0,1)$ & $(1,0)$ & $(0,0)$ & $(0,1)$ & $(1,0)$ & $(0,0)$ &
$(0,1)$ & $(1,0)$ & $(0,0)$ & $(0,1)$ & $(1,0)$\\\hline\hline
$0.6$ & $0.79091$ & $0.79092$ & $0.79825$ & $0.74058$ & $0.74072$ & $0.76376$
& $0.69493$ & $0.69515$ & $0.71279$ & $0.67934$ & $0.67942$ & $0.68707$%
\\\hline
$0.7$ & $0.68601$ & $0.68602$ & $0.70800$ & $0.56714$ & $0.56724$ & $0.62744$
& $0.42764$ & $0.42845$ & $0.49171$ & $0.32352$ & $0.32392$ & $0.37485$%
\\\hline
$0.8$ & $0.55636$ & $0.55681$ & $0.59157$ & $0.36791$ & $0.36792$ & $0.46662$
& $0.16655$ & $0.16661$ & $0.25701$ & $0.05710$ & $0.05713$ & $0.09801$%
\\\hline
$0.9$ & $0.40163$ & $0.40471$ & $0.43235$ & $0.21187$ & $0.21231$ & $0.30477$
& $0.07522$ & $0.07532$ & $0.12658$ & $0.02413$ & $0.02416$ & $0.04180$%
\\\hline
\end{tabular}
\caption{Black hole charge $q$ and spin $\chi$ of representative clouds on the
existence lines of fundamental and excited modes for $\alpha=-10^{2}$,
$-10^{3}$, $-10^{4}$ and $-10^{5}$. For a given $\alpha$ and $\chi$,
fundamental clouds with $\left(  n,l\right)  =\left(  0,0\right)  $ require
smaller $q$ than excited clouds, indicating that they are more prone to
formation through the tachyonic instability.}%
\label{tab:qanx}%
\end{table}

Beyond the fundamental mode, excited modes of scalar clouds can also form
around KN black holes, potentially leading to excited states of scalarized
black holes. Fig. \ref{fig:ExistenceDomainExitedMode} illustrates existence
lines in the $\left(  \chi,q\right)  $ space for excited scalar clouds with
$\left(  n,l\right)  =\left(  1,0\right)  $ and $\left(  0,1\right)  $, which
exhibit similarities to the fundamental mode. Specifically, the existence
lines of excited clouds lie between the extremal and critical existence lines,
with both endpoints resting on the extremal line. As $\alpha$ approaches
$\alpha_{\text{cr}}$ from below, the existence lines contract and converge to
the critical point $B$, indicating excited clouds cannot form for
$\alpha>\alpha_{\text{cr}}$. Moreover, as $\alpha$ becomes more negative, the
existence lines shift closer to the critical existence line. To compare the
existence lines of fundamental and excited modes, we provide the $q$ and
$\chi$ values of representative clouds for various $\alpha$ in Tab.
\ref{tab:qanx}. It is evident that, for a given $\alpha$ and $\chi$, KN black
holes require the smallest $q$ to support fundamental clouds, indicating that
a stronger tachyonic instability is needed to form excited clouds.
Additionally, the existence line of the $\left(  n,l\right)  =\left(
0,1\right)  $ excited mode lies just slightly above that of the fundamental
mode, suggesting that $n=0$ scalar clouds are more easily generated than those
with $n=1$.

\begin{figure}[ptb]
\begin{centering}
\includegraphics[scale=0.59]{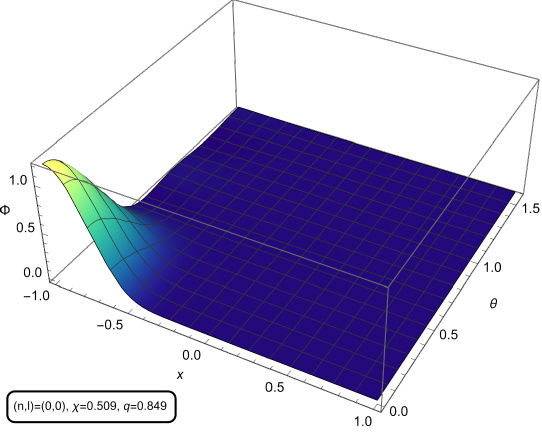} \includegraphics[scale=0.59]{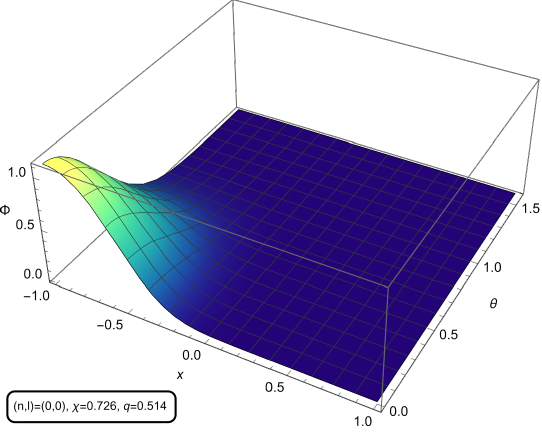}
\includegraphics[scale=0.59]{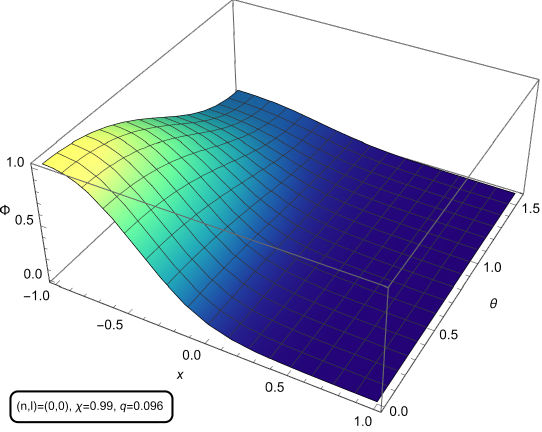}
\[
\quad
\]
\includegraphics[scale=0.59]{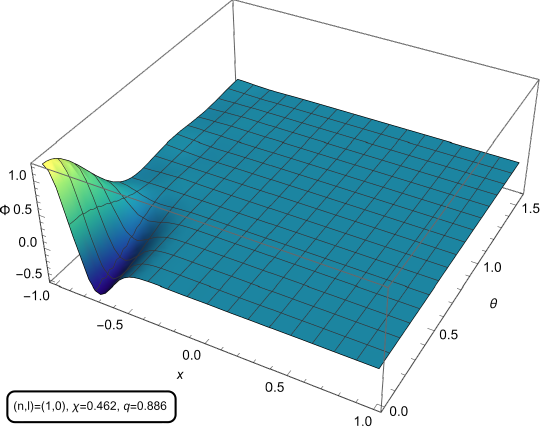} \includegraphics[scale=0.59]{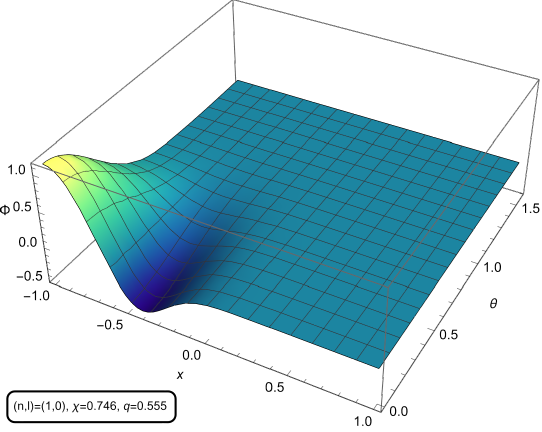}
\includegraphics[scale=0.59]{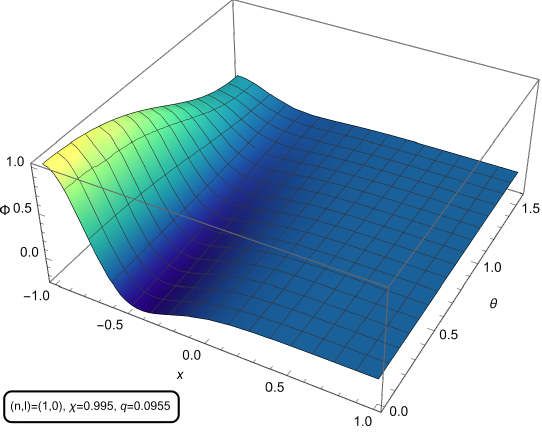}
\[
\quad
\]
\includegraphics[scale=0.59]{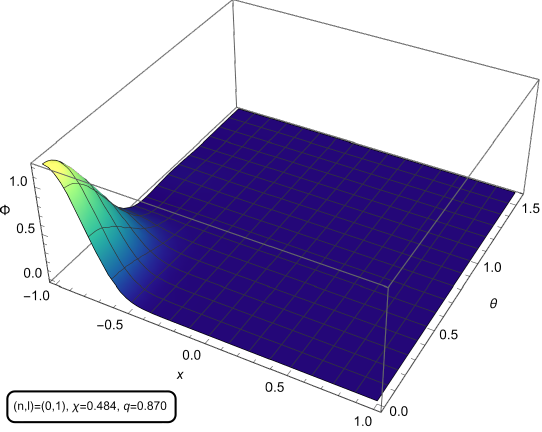} \includegraphics[scale=0.59]{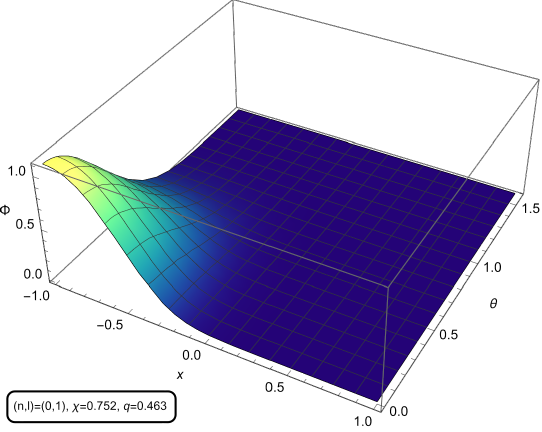}
\includegraphics[scale=0.59]{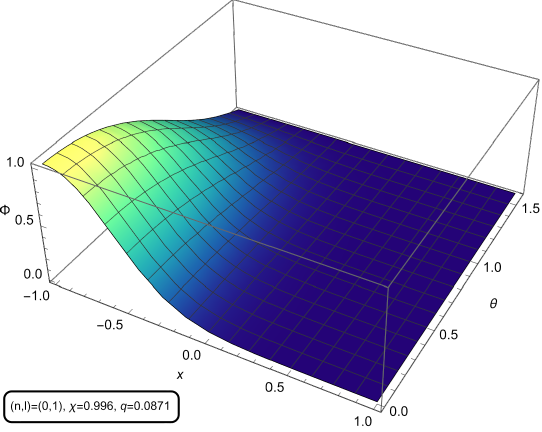}
\par\end{centering}
\caption{Wave function $\phi\left(  x,\theta\right)  $ of representative
scalar clouds for $\left(  n,l\right)  =\left(  0,0\right)  $ (\textbf{Top
Row}), $\left(  n,l\right)  =\left(  1,0\right)  $ (\textbf{Middle Row}) and
$\left(  n,l\right)  =\left(  0,1\right)  $ (\textbf{Bottom Row}) with
$\alpha=-10^{3}$. For all cases, we set $\phi\left(  x,\theta\right)  =1$ at
$\left(  x,\theta\right)  =\left(  -1,0\right)  $. Scalar cloud wave functions
are concentrated near the black hole's poles, while black hole rotation has a
tendency to spread wave functions towards the equatorial plane.}%
\label{fig:waveform}%
\end{figure}

Finally, we present representative scalar cloud wave functions for $\left(
n,l\right)  =\left(  0,0\right)  $, $\left(  1,0\right)  $ and $\left(
0,1\right)  $ in Fig. \ref{fig:waveform}. Selecting three cloud solutions on
each existence line with $\alpha=-10^{3}$, all wave functions exhibit
concentrations near the event horizon and the poles. For a fixed $r$ close to
the event horizon, the wave functions gradually decrease along the $\theta$
direction, reaching a minimum at the equatorial plane. As the black hole's
spin increases, the concentration of wave functions tends to spread towards
the equatorial plane. It is noteworthy that rapidly rotating black holes with
$\alpha>0$ display scalar cloud concentrations near the equatorial plane
\cite{Guo:2024bkw}. Beyond these commonalities, $\left(  n,l\right)  =\left(
1,0\right)  $ scalar clouds feature a radial node, resulting in a valley along
the $\theta$ direction within their wave functions. This valley approaches the
event horizon as the spin increases. For $\left(  n,l\right)  =\left(
0,1\right)  $ scalar clouds, their odd parity with respect to the equatorial
plane causes their wave functions to vanish at $\theta=\pi/2$.

\section{Conclusions}

\label{Sec:Conc}

In this paper, we have explored scalar clouds generated by the spin-induced
tachyonic instability around KN black holes within the framework of the EMS
model, focusing on both the fundamental and excited modes. By employing the
spectral method, we have successfully identified the parameter space where
such scalar clouds can exist. Our findings reveal that the existence of scalar
clouds is contingent upon the interplay between the black hole's charge, spin
and the coupling constant $\alpha$.

Specifically, we have determined that for a given $\alpha$, there exists a
distinct existence line in the $\left(  \chi,q\right)  $ parameter space along
which scalar clouds can form. Notably, this existence line intersects the
extremal line at both endpoints, implying that the tachyonic instability is
insufficient to induce scalar cloud formation for black holes that rotate
either too slowly or too rapidly. Additionally, our analysis reveals that the
region of the parameter space where scalar clouds exist shrinks as $\alpha$
approaches a critical value, $\alpha_{\text{cr}}\simeq-13.398$. This
observation suggests that scalar clouds cannot form for $\alpha$ values
greater than $\alpha_{\text{cr}}$, a conclusion further supported by the
existence domains presented in the $\left(  \alpha,\chi\right)  $ and $\left(
\alpha,q\right)  $ planes.

Previous studies \cite{Hod:2022txa,Lai:2022ppn} have established constraints
on the existence domain of scalar clouds, as expressed in Eqs. $\left(
\ref{eq:constx}\right)  $ and $\left(  \ref{eq:consxq}\right)  $. These
constraints were suggested to be saturated in the strong coupling limit
($\alpha\rightarrow-\infty$). Our numerical results corroborate these
findings. Furthermore, we also showed that for $\chi>1/\sqrt{2}$, a portion of
the existence lines converge towards $q=0$ in the strong coupling limit,
suggesting that the formation of scalar clouds requires only a minimal amount
of charge.

Our investigation studies the influence of the scalar field mode on scalar
cloud formation. While the fundamental mode requires the least charge for
formation, excited modes necessitate a stronger tachyonic instability.
Additionally, we have observed that scalar cloud wave functions are
concentrated near the black hole's poles, differing from the concentration
near the equatorial plane in the $\alpha>0$ case. As the black hole's spin
increases, the concentration of scalar clouds near the poles becomes less
pronounced. This wave function behavior is consistent across different modes,
although excited scalar clouds exhibit additional features such as radial
nodes or odd parity with respect to the equatorial plane.

Since scalar clouds mark the onset of scalarization from scalar-free black
holes, the findings presented in this study provide a foundation for future
research on non-linear realizations of scalar clouds, namely spin-induced
scalarized KN black holes. These explorations may contribute to a deeper
understanding of spontaneous scalarization. Additionally, future research
could delve into the non-linear dynamics of scalar clouds and their potential
implications for black hole stability and related astrophysical phenomena.

\begin{acknowledgments}
We are grateful to Yiqian Chen for useful discussions and valuable comments.
This work is supported in part by NSFC (Grant Nos. 12105191, 12275183,
12275184, 11875196, 12347133 and 12250410250).
\end{acknowledgments}

\appendix

\section*{Appendix: Convergence Test}

\label{Sec:app}

\begin{figure}[ptb]
\begin{centering}
\includegraphics[scale=0.6]{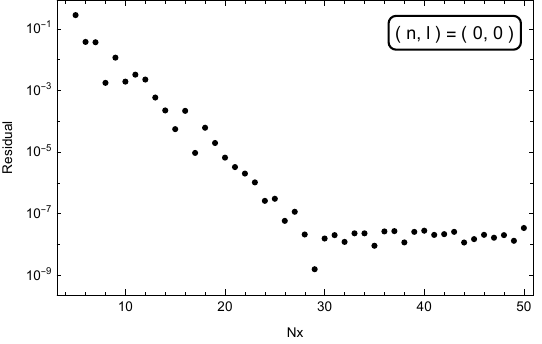}\includegraphics[scale=0.6]{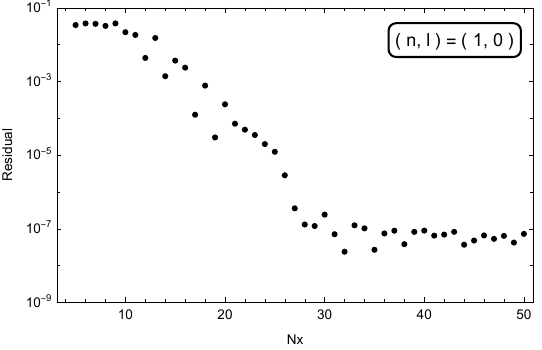}\includegraphics[scale=0.6]{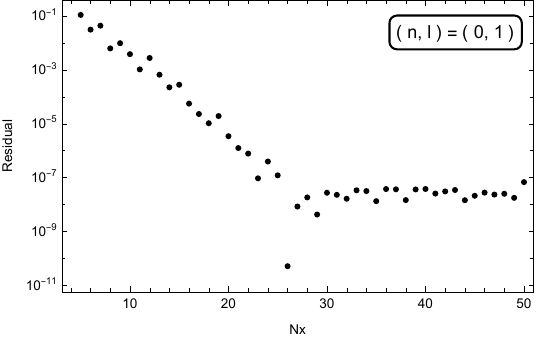}
\par\end{centering}
\begin{centering}
\includegraphics[scale=0.6]{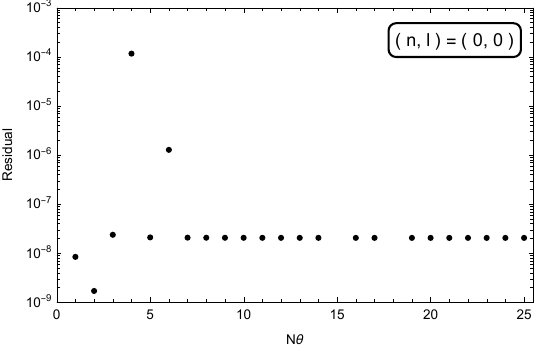}\includegraphics[scale=0.6]{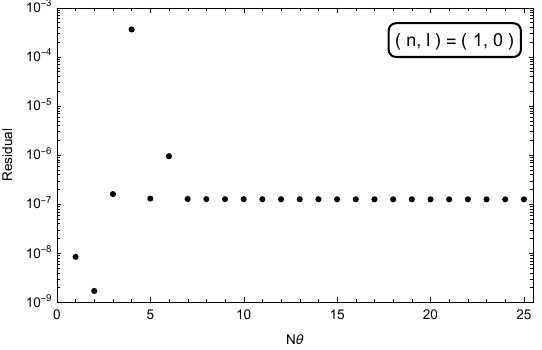}\includegraphics[scale=0.6]{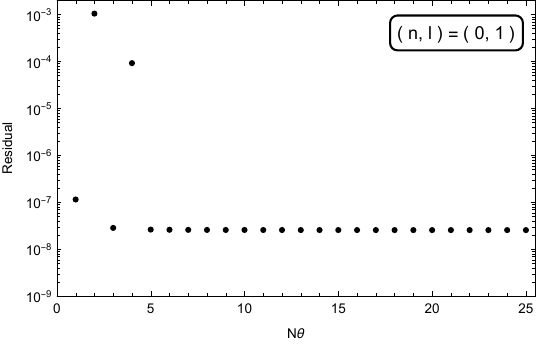}
\par\end{centering}
\caption{Logarithmic plot of the residual error as a function of $N_{x}$
(\textbf{Top Row}) and $N_{\theta}$ (\textbf{Bottom Row}) for scalar clouds
with $\left(  n,l\right)  =\left(  0,0\right)  $, $\left(  1,0\right)  $ and
$\left(  0,1\right)  $. In the top row, $N_{\theta}=5$ is fixed, while in the
bottom row, $N_{x}=28$ is held constant. All scalar cloud solutions share the
same $\alpha$ and $a/r_{+}^{2}$, namely $\alpha=-10^{3}$ and $a/r_{+}^{2}%
=0.8$. Exponential convergence is evident, with a round-off plateau observed.}%
\label{fig:Convergence Test}%
\end{figure}

In this appendix, we assess the convergence of our numerical code by
calculating fundamental and excited scalar cloud solutions with $\alpha
=-10^{3}$ and $a/r_{+}^{2}=0.8$ at various resolutions. The top row of Fig.
\ref{fig:Convergence Test} depicts the maximum absolute value of the residual
error as a function of the radial resolution $N_{x}$ with $N_{\theta}=5$. All
scalar cloud solutions demonstrate exponential convergence, with a round-off
plateau approximately at $N_{x}\geq30$. The bottom row presents the maximum
absolute value of the residual error as a function of the angular resolution
$N_{\theta}$ with $N_{x}=28$. While some outliers occur at low $\theta$
resolutions, exponential convergence is observed overall, with a convergence
plateau reached for $N_{\theta}\geq5$. To maintain a residual tolerance of
$10^{-7}$, we adopt $\left(  N_{x},N_{\theta}\right)  =\left(  28,5\right)  $
in our numerical calculations.

\bibliographystyle{unsrturl}
\bibliography{ref}

\end{document}